\documentclass[amsmath,amssymb,aps,nofootinbib,showkeys,showpacs]{elsarticle}
\usepackage[utf8]{inputenc}
\usepackage{graphicx}
\usepackage{comment}
\usepackage{natbib}
\usepackage{enumerate}
\usepackage{amsmath}
\usepackage{amssymb}
\usepackage{mathrsfs}
\usepackage{lineno, comment, url}
\usepackage[usenames,dvipsnames,svgnames,table]{xcolor}

\usepackage[bbgreekl]{mathbbol}
\usepackage{units}

\usepackage[english]{babel}

\newcommand{\rr}{\mathbf{r}}
\newcommand{\pp}{\mathbf{p}}
\newcommand{\rrp}{\mathbf{r}'}
\newcommand{\ppp}{\mathbf{p}'}
\newcommand{\FF}{\mathbf{F}}
\newcommand{\const}{\mathrm{const}}
\newcommand{\eps}{\varepsilon}
\newcommand{\R}{\mathbb{R}}
\newcommand{\OO}{\mathcal{O}}
\newcommand{\xx}{\mathbf{x}}
\newcommand{\yy}{\mathbf{y}}
\newcommand{\proj}{\mathbf{\pi}}
\newcommand{\JJ}{\mathbf{J}}

\newcommand{\diff}{\mathrm{d}\,}

\definecolor{MyBlue}{rgb}{0.,0.1,1}

\newcommand{\kk}{\mathbf{k}}
\newcommand{\uu}{\mathbf{u}}
\newcommand{\barn}{\bar{\rho}}
\newcommand{\tilden}{\tilde{\rho}}
\newcommand{\dL}{{^{\downarrow}L}}
\newcommand{\dM}{{^{\downarrow}M}}
\newcommand{\MM}{\mathbf{M}}
\newcommand{\dMM}{{^{\downarrow}\MM}}
\newcommand{\dJ}{{^{\downarrow}J}}
\newcommand{\LL}{\mathbf{L}}

\newcommand{\dE}{{^\downarrow}E}
\newcommand{\dS}{{^\downarrow}S}

\begin{document}

\begin{frontmatter}
\title{Statistical mechanics of Landau damping}

\author[mff]{Michal Pavelka\corref{cor1}}
\ead{pavelka@karlin.mff.cuni.cz}
\author[FJFI]{V\'{a}clav Klika}
\ead{vaclav.klika@fjfi.cvut.cz}
\author[Poly]{Miroslav Grmela}
\ead{miroslav.grmela@polymtl.ca}

\address[mff]{Mathematical Institute, Faculty of Mathematics and Physics, Charles University in Prague, Sokolovská 83, 186 75 Prague, Czech Republic}
\address[FJFI]{Department of Mathematics, FNSPE, Czech Technical University in Prague, Trojanova 13, 120 00 Prague, Czech Republic}
\address[Poly]{\'{E}cole Polytechnique de Montr\'{e}al, C.P.6079 suc. Centre-ville, Montréal, H3C 3A7, Québec, Canada}
\cortext[cor1]{Corresponding author}
\journal{Physica A: Statistical mechanics and its applications}

\begin{abstract}
Landau damping is the tendency of solutions to the Vlasov equation towards spatially homogeneous distribution functions. The distribution functions however approach the spatially homogeneous manifold only weakly, and Boltzmann entropy is not changed by Vlasov equation. On the other hand, density and  kinetic energy density, which are integrals of the distribution function, approach spatially homogeneous states strongly, which is accompanied by growth of the hydrodynamic entropy. Such a behavior can be seen when Vlasov equation is reduced to the evolution equations for density and  kinetic energy density by means of the Ehrenfest reduction.
\end{abstract}
\end{frontmatter}


\numberwithin{equation}{section}


\section{Introduction}
\label{sec.introduction}
Dynamics of self-interacting gas is well described by Vlasov equation, 
\begin{equation}\label{eq.Vlasov}
\frac{\partial f(t, \rr, \pp)}{\partial t} = - \frac{\pp}{m} \cdot \frac{\partial f(t,\rr,\pp)}{\partial \rr} - \FF(f(t,\rr,\pp))\cdot \frac{\partial f}{\partial \pp},
\end{equation}
where $m$ is mass of one particle and $f(t, \rr, \pp)$ is the one-particle distribution function on phase space and $\FF$ is the force exerted on the gas from outside or by the gas itself. The force can be derived from a Hamiltonian (or energy of the system) as follows.\footnote{For simplicity of notation, the dependence of the distribution function on time or on position and momentum will not always be written explicitly.} Energy of the system with a self-consistent long-range interaction is
\begin{equation}\label{eq.energy}
E = \int d\rr \int d\pp \frac{\pp^2}{2m} f + \frac{1}{2}\int d\rr \int d\pp \int d\rrp\int d\ppp f(\rr,\pp) V(|\rr-\rrp|)  f(\rr',\pp'),
\end{equation}
where $V(|\rr-\rrp|)$ is an interaction potential, e.g. electrostatic or gravitational potential.
Derivative of energy with respect to the distribution functions is then
\begin{equation}
\frac{\delta E}{\delta f(\rr,\pp)} = \frac{\pp^2}{2m} +  \int d\rrp\int d\ppp V(|\rr-\rrp|)  f(\rr',\pp').
\end{equation}
The force is then given by
\begin{equation}
\FF(t, \rr) = -\int d\rrp\int d\ppp \frac{\partial V(|\rr-\rrp|)}{\partial \rr}  f(\rr',\pp').
\end{equation}
Note that this dependence of the force on the distribution function makes Vlasov equation nonlinear. 
All terms in the Vlasov equation \eqref{eq.Vlasov} have now been specified. 

Landau damping is a feature of  solutions of Vlasov equation -  solutions approach spatially homogeneous distributions.  This was first derived by L. D. Landau \cite{Landau-damping,Landau10} when considering linearized Vlasov equation. C. Villani and C. Mouhot then proved  it without the linearization, i.e. in the fully nonlinear setting (see \cite{Mouhot2011,Villani2014}, where conditions on the initial perturbation and decay rates can be found). More  precisely, Landau damping can be seen as a development of fast oscillation in the $\pp-$dependence of $f(t, \rr,\pp)$ and a simultaneous decay of the dependence on $\rr$. The distribution function approaches a spatially homogeneous, i.e. $\rr-$ independent, distribution function only weakly due to the fast oscillations in the $\pp-$ dependence. However, the approach of integral over the $\pp-$ coordinate (particle density $\rho(\rr)$) is already strong, i.e.
\begin{equation} f(\rr,\pp)\stackrel{w}{\rightarrow} \tilde{f}(\pp) \qquad\mbox{and}\qquad \rho(\rr) = \int d\pp m f(\rr,\pp) \rightarrow \tilde{\rho} =\const.
\end{equation}
Constant $m$ is the mass of one particle.
These two properties of solutions to the Vlasov equation are referred to as the Landau damping. 

Villani \cite{Villani2014} summarizes the results of the analytic computations in the linear\footnote{The second term on the r.h.s. of Eq. \eqref{eq.Vlasov} is linearized.} Landau damping as follows. Under certain assumptions (the data are analytic and the Penrose stability condition holds),  solutions to the linearized Vlasov equation  have the following properties: (i) all nonzero spatial Fourier modes of $\rho^1(t,\rr)= \int h(t,\rr,\pp) d\pp$ ($h$ stands for the perturbation of a spatially homogeneous $f_0$ distribution function), converge to 0, hence $\rho^1$ converges to a constant in space; (ii) the force converges to 0 also; (iii) $\int h(t,\rr,\pp) \varphi(\rr,\pp) d\rr d\pp$ converges to $\int h_i(\rr',\pp)\pp\varphi(\rr,\pp) d\rr d\rr' d\pp$ for any smooth test function $\varphi$ (where $h_i(\rr,\pp)=h(0,\rr,\pp)$ is the initial value of the perturbation).

The nonlinear Landau damping  enjoys the same long-time properties if the initial perturbation is a perturbation of a stable equilibrium and some other assumptions are satisfied. Particularly that the interaction potential be no more singular than Newtonian (Fourier transpose $\hat{W}(\kk)=O(1/|\kk|^2)$ for $\kk\in Z^d$). The main two observations are again that (i) $f(t,\rr,\pp)\rightarrow f_{+\infty} (\pp)$ weakly, which implies convergence of smooth observables, that is $\int f(t,\rr,\pp) \varphi(\rr,\pp) d\rr d\pp \rightarrow \int f_{+\infty} (\pp) \varphi(\rr',\pp) d\rr d\rr' d\pp = V \int f_{+\infty} (\pp) \varphi(\rr',\pp) d\rr' d\pp$), and $F(t,\rr)\rightarrow 0$ strongly with time.

Landau damping is also interesting from the thermodynamic point of view. Although it clearly described decay of the distribution function to homogeneous equilibria, the Boltzmann entropy,
\begin{equation}\label{eq.S.B}
S^{(B)}(f) = -k_B \int d\rr \int d\pp f(\rr,\pp) \left(\ln\left(h^3 f(\rr,\pp)\right)-1\right),
\end{equation}
is kept constant by Vlasov equation, as can be verified by direct calculation. This seems to be in contradiction with the usual thermodynamic interpretation of relaxation to some equilibrium, which is usually associated with growth of entropy. This seemingly paradoxical situation will be resolved in this paper by showing that the hydrodynamic entropy, which is more macroscopic than the Boltzmann entropy, grows while Boltzmann entropy is conserved. This seems to be in agreement with the point of view of C. Villani \cite{Villani-Praha}.

In this paper we follow \cite{KRM} and investigate the Vlasov equation (I.1) by applying methods of statistical mechanics. This investigation offers  us a possibility to see both the statistical methods and the Landau damping in a new perspective. In \cite{KRM} we have first extended the Vlasov equation to include explicitly the micro-turbulence and then we have shown that its decay brings about the Landau damping. In this paper we apply the methods of non-equilibrium statistical mechanics, namely the Maximum Entropy Principle (MaxEnt) and the Ehrenfest reduction method (see more in Section II) directly to the Vlasov equation. We reduce it to two equations governing the time evolution of two scalar fields, $\rho(\rr)=\int d\pp m f(\rr,\pp)$ and $\epsilon(\rr)=\int d\pp \frac{\pp^2}{2m}f(\rr,\pp)$. From an analysis of their solutions we are then able to see the Landau damping as the approach of the field $\rho(\rr)$ to a spatially homogeneous field (in Section III).

\section{Ehrenfest reduction}
A reduction of a dynamical system (DS1) to another dynamical system (DS2) is a pattern recognition in the phase portrait corresponding to (DS1). The phase portrait corresponding to (DS1) is a collection of the trajectories  that arise in (DS1) (i.e. solutions to the governing equations of (DS1)). The recognized pattern in the  (DS1) phase portrait is the phase portrait corresponding to (DS2). The first step in the reduction is thus an information about the phase portrait.  How shall we obtain such information? We recall two examples.

First, it is the Gibbs reduction of the Liouville equation to the equilibrium thermodynamics. The phase portrait is assumed to be ergodic and the recognized pattern is the Gibbs equilibrium distribution function obtained by maximizing the Gibbs entropy with  constraints (MaxEnt principle). In the Gibbs analysis the constraints are the state variables of the equilibrium thermodynamics (i.e. the total mass and the total energy). The Gibbs entropy is a measure of disorder. Since the constraints remain constant during the time evolution, the reduced time evolution is no time evolution.

The second example is the Ehrenfest reduction of (DS1) to (DS2). The constraints are now fields that serve as state variables in (DS2). These two fields  do not remain constant in the (DS1) time evolution. The pattern is again obtained by the MaxEnt principle and its time evolution by following the (DS1) time evolution in a small time interval. Below, we describe the Ehrenfest reduction in detail (in the rest of this section) and then we apply it on the Vlasov equation (I.1) in Section III.

\subsection{General formulation}
Let us first recall the Ehrenfest reduction, that was developed in \cite{GK-Ehrenfest,GK-Ehrenfest2}. The starting point is to expand solutions to the  (DS1) time evolution equation (denoted by state variables $\xx$),
\begin{equation}\label{eq.evo.x}
\dot{\xx} = J(\xx)
\end{equation}
in Taylor series in time (i.e. close to the initial condition $\xx(t+\tau)|_{\tau=0}=\xx(t)$),
\begin{equation}
\xx(t + \tau) = \xx(t) + \tau \dot{x}|_t + \frac{\tau^2}{2} \ddot{\xx}|_t + \OO(\tau^3).
\end{equation}
By the symbol $\xx$ we denote the state variables of the (DS1) dynamics.
Substituting Eq. \eqref{eq.evo.x} into the expansion leads to 
\begin{equation}\label{eq.x.Taylor}
\xx(t + \tau) = \xx(t) + \tau J(\xx(t)) + \frac{\tau^2}{2} \left\langle\frac{\delta J(\xx(t))}{\delta \xx}, J(\xx(t))\right\rangle + \OO(\tau^3).
\end{equation}

A less detailed level of description, i.e. (DS2), has state variables
\begin{equation}
\yy = \langle \proj, \xx\rangle,
\end{equation}
where $\langle\proj,\bullet\rangle$ is the 
projection operator, derivatives of which $\frac{\partial \pi^a}{\partial x^j}$ will be denoted simply as $\pi^a_i$. For linear constant-in-time projection operators the derivatives constitute a constant matrix. 

Exact evolution of variables $\yy$ can be obtained by
\begin{equation}
\dot{\yy} = \langle\proj, \dot{x}\rangle = \langle\proj, J(\xx)\rangle, 
\end{equation}
but in order to evaluate it, it would be necessary to solve equation \eqref{eq.evo.x}. Since the aim is to obtain evolution equation for $\yy$ only in terms of $\yy$, a reduction has to be carried out. In other words, some information has to be forgotten. The approximation can be sought by following two routes, i) projecting the more detailed evolution to the less detailed state variables; ii) expanding the sought evolution equation on the coarser level. Comparison of these two approaches yields an assessment of Taylor series expansion of the sought evolution equation on the coarser level (to arbitrary order). 

In the latter route we search for the unknown operator $\phi$ that governs the evolution on the coarser level  (DS2)
$$ \dot{\yy} = \phi(\yy) .$$
Expanding its solution around an arbitrary initial condition $\yy(t)$ yields
\begin{equation} \label{eq.LowerLevel_expansion}
y_k(t+\tau) = y_k(t)+\tau\phi_k(\yy(t))+\frac{\tau^2}{2} \frac{\delta \phi_k}{\delta y_j} \phi_j.
\end{equation}
To identify the asymptotic expansion of the unknown operator $\phi$, we write
\begin{equation} \label{eq.phi_expansion}
	\phi_k=R_k^{(0)}+\tau R_k^{(1)}+\frac{\tau^2}{2}R_k^{(2)}+O(\tau^3).
\end{equation}

Now we compare this coarser (DS2) level expansion  to the (DS1) time evolution. In order to achieve that, however, the  (DS1) time evolution, has to be made dependent on $\yy$.  We choose the mapping from $\yy$ to $\xx$  to be the MaxEnt mapping $\tilde{\xx}(\yy)$, i.e.  to find $\xx$ such that the detailed entropy $S(\xx)$ is maximal subject to the constraints represented by $\yy=\proj(\xx)$. Then, by substituting \eqref{eq.phi_expansion} into \eqref{eq.LowerLevel_expansion}) with projection of Eq. \eqref{eq.x.Taylor},  we obtain
\begin{equation}\label{eq.x.Taylor.proj}
\langle\pi,\xx(t + \tau)\rangle = 
 \langle\pi,\xx(t)\rangle + \tau \langle\pi,J(\tilde\xx(\yy))\rangle + \frac{\tau^2}{2} \left\langle\pi,\left\langle\frac{\delta J(\xx(t))}{\delta \xx}\Bigg|_{\tilde\xx(\yy)}, J(\tilde\xx(\yy))\right\rangle\right\rangle + \OO(\tau^3),
\end{equation} 
where the quasi-equilibrium $\tilde\xx(\yy)$ initial conditions were chosen so that only $\yy$ appears in the expansion.
  The comparison leads to
\begin{subequations}
\begin{align}
y_k&=\langle\pi_k,\tilde\xx(\yy)\rangle \text{ $\ldots$from $\tau^0$ coefficient,}\\
\label{eq.R0} R_k^{(0)} &= \langle\pi_k,J(\tilde{\xx}(\yy))\rangle\text{ $\ldots$ from $\tau^1$},
\end{align}
and
\begin{align}\label{eq.R1}
R_k^{(1)} &= \frac{1}{2}\left( \left\langle\pi_k,\left\langle\frac{\delta J(\xx(t))}{\delta \xx}\Bigg|_{\tilde\xx(\yy)}, J(\tilde\xx(\yy))\right\rangle\right\rangle  -
\left\langle \frac{\delta R_k^{(0)}}{\delta y_j}, R_j^{(0)}\right\rangle \right)\text{ $\ldots$ from $\tau^2$}\nonumber\\
&=\frac{1}{2}\left( \left\langle\pi_k,\left\langle\frac{\delta J(\xx(t))}{\delta \xx}\Bigg|_{\tilde\xx(\yy)}, J(\tilde\xx(\yy))\right\rangle\right\rangle  -
\left\langle \frac{\delta \langle\pi_k,J(\tilde{\xx}(\yy))\rangle}{\delta y_j}, \langle\pi_j,J(\tilde{\xx}(\yy))\rangle\right\rangle \right)
\end{align}
\end{subequations}
which completes the estimate of the sought evolution equation on the coarser level via \eqref{eq.phi_expansion}
\begin{equation}\label{eq.evo.y.final}
\dot y_k=\phi(\yy)\approx R_k^{(0)}+\tau R_k^{(1)} +O(\tau^2).
\end{equation}

Note that the $\tau^0$ term has to be strictly satisfied due to the requirement that both the more and less detailed evolution coincide at $\tau=0$ on the quasi-equilibrium. The zeroth correction $R_k^{(0)}$ can be also naturally understood as the least biased way to express $\xx$ in terms of $\yy$ is the MaxEnt mapping from $\yy$ to $\xx$, $\tilde{\xx}(\yy)$, which is exactly what the first approximation is. 
The first correction $R_k^{(1)}$ is typically non-zero as the detailed evolution \eqref{eq.evo.x} carries $\xx$ from the quasi-equilibrium $\tilde{\xx}(\yy)$ to values which are not in the image of the MaxEnt mapping, i.e. out of the quasi-equilibrium (or Legendre) submanifold, and corrections have to be introduced.


\subsection{Hamiltonian version of Ehrenfest reduction}
The purpose of this section is to reformulate the Ehrenfest reduction in the case when the detailed evolution (DS1) is Hamiltonian.

\subsubsection{Hamiltonian structure of Vlasov equation}
Vlasov equation \eqref{eq.Vlasov} is a Hamiltonian evolution, since it is generated by the Boltzmann-Poisson bracket
\begin{equation}
\{A,B\}^{(B)} = \int d\rr\int d\pp f \left(
\frac{\partial}{\partial \rr}\frac{\delta A}{\delta f}\frac{\partial}{\partial \pp}\frac{\delta B}{\delta f}
-\frac{\partial}{\partial \rr}\frac{\delta B}{\delta f}\frac{\partial}{\partial \pp}\frac{\delta A}{\delta f}\right),
\end{equation}
which is of course antisymmetric and fulfills both Leibniz rule and Jacobi identity.
By rewriting the bracket as 
\begin{equation}
\int d\rr \int d\pp \frac{\delta A}{\delta f} \cdot \frac{\partial f}{\partial t}=
\{A,E\}^{(B)} = \int d\rr \int d\pp \frac{\delta A}{\delta f} \cdot \mathrm{rhs},
\end{equation}
the terms $\mathrm{rhs}$ are the right hand side of the evolution equation for $f$, which is Vlasov equation \eqref{eq.Vlasov}. This Poisson bracket can be derived from the Liouville Poisson bracket by projection or it can be seen as the Lie-Poisson bracket corresponding to symplectic transformations on the one-particle cotangent bundle, see e.g. \cite{PhysicaD-2015}.

Any real-valued function of the distribution function $\sigma:\R\rightarrow\R$ then generates a Casimir of the bracket, 
\begin{equation}
\left\{A, \int d\rr\int d\pp \sigma(f)\right\}^{(B)} = 0 \qquad \forall A(f).
\end{equation}
This means in particular that 
\begin{equation}
\dot{S}^{(B)} = \left\{S^{(B)}, E\right\} = 0,
\end{equation}
and Boltzmann entropy is thus indeed conserved by the Vlasov equation. The Boltzmann entropy is a plausible entropy (a Casimir of the Boltzmann Poisson bracket), but it is not the only one possible. Nevertheless, in the case of ideal gases it follows by MaxEnt from the Liouville entropy, see e.g. \cite{RedExt}, which can be seen as the phase-space analogue of the Shannon entropy $-k_B \sum_i p_i \ln p_i$. Shannon entropy is uniquely determined by Shannon's axioms defining uncertainty \cite{Shannon}. In this sense the Boltzmann entropy $S^{(B)}$ is unique.
Moreover, energy is conserved automatically due to the antisymmetry of the Poisson bracket.

Each Poisson bracket can be seen as constructed by means of a Poisson bivector $\LL$, 
\begin{equation}
\{A,B\} = A_{x^i} L^{ij} B_{x^j}.
\end{equation}
Antisymmetry of the Poisson bracket is expressed by antisymmetry of the Poisson bivector, and Jacobi identity for the bracket is also inherited from a corresponding formula for the bivector, see e.g. \cite{Fecko}, p. 332.

\subsubsection{Formal solution of Hamiltonian evolution}
The first step in Ehrenfest reduction was to expand the evolution of detailed variables $\xx$ as a Taylor series in time. Assume that evolution of $\xx$ is purely Hamiltonian, which means that
\begin{equation}\label{Ham:eq.x.evo}
\dot{x}^i = \underbrace{L^{ij} E_{x^j}}_{=J^i} \qquad \mbox{or} \qquad \dot{A}(\xx) = \{A,E\}\quad\forall A.
\end{equation}
Solution to the Hamiltonian evolution equations can be expanded in series as (see e.g. \cite{Fecko}, pp. 334)
\begin{equation}
A(\xx(t+\tau)) = A(\xx(t)) + \tau \{A,E\} + \frac{\tau^2}{2} \{\{A,E\},E\} + \OO(\tau^3).
\end{equation}
When expanding the Poisson brackets by means of the Poisson bivector and when taking into account only linear functionals $A$, i.e. $A(\xx) = A_i x^i$, this last equation becomes
\begin{equation}\label{Ham:eq.x.Taylor}
A_i x^i(t + \tau) = A_i x^i(t) + \tau A_i L^{ij}E_{x^j} + \frac{\tau^2}{2} A_i\frac{\partial}{\partial x^k}\left(L^{ij}E_{x^j}\right)L^{kl}E_{x^l}  + \OO(\tau^3) \quad \forall A_i,
\end{equation}
which is analogical to Eq. \eqref{eq.x.Taylor}. 

Taylor expansion \eqref{Ham:eq.x.Taylor} can be interpreted as a new evolution equation for $x^i$
\begin{equation}\label{eq.x.reg}
\frac{d x^i}{d \tau} =  L^{ij}E_{x^j} + \frac{\tau}{2} \frac{\partial}{\partial x^k}\left(L^{ij}E_{x^j}\right)L^{kl}E_{x^l},
\end{equation}
where the first term on the right hand side generates reversible evolution while the second irreversible. Such an evolution equation can be regarded as a \textbf{self-regularized} evolution of $x^i$ corresponding to time step $\tau$. Similar idea has already been brought up in symplectic integration \cite{Leimkuhler}.

The original Hamiltonian evolution, Eq. \eqref{Ham:eq.x.evo}, conserved both energy (due to antisymmetry) and entropy (due to degeneracy of the Poisson bracket). None of these properties is kept in the self-regularized evolution equations \eqref{eq.x.reg}. Indeed, evolution of entropy becomes
\begin{subequations}
\begin{eqnarray}
\frac{d S}{d \tau} &=& S_{x^i} \frac{d x^i}{d \tau}=  \underbrace{S_{x^i}L^{ij}}_{=0}E_{x^j} + \frac{\tau}{2} S_{x^i} \frac{\partial}{\partial x^k}\left(L^{ij}E_{x^j}\right)L^{kl}E_{x^l}\nonumber\\
&=& \frac{\tau}{2} \frac{\partial}{\partial x^k}\left(\underbrace{S_{x^i} L^{ij}}_{=0}E_{x^j}\right)L^{kl}E_{x^l}
-\frac{\tau}{2} S_{x^k x^i} L^{ij}E_{x^j}L^{kl}E_{x^l}\nonumber\\
&=&\frac{\tau}{2} L^{ij}E_{x^j}\left(-S_{x^i x^k}\right) L^{kl}E_{x^l}\geq 0.
\end{eqnarray}
The inequality follows from concavity of $S(\xx)$ (and thus negative semi-definiteness of $d^2 S$). Entropy is thus produced by the self-regularized equations. Similarly, evolution of energy becomes
\begin{equation}
\frac{d E}{d \tau} = E_{x^i} \frac{d x^i}{d \tau} = 
-\frac{\tau}{2} L^{ij}E_{x^j}E_{x^i x^k} L^{kl}E_{x^l}\leq 0
\end{equation}
due to convexity of energy ($d^2 E$ positive semi-definite). Energy is thus dissipated by the self-regularized evolution.
\end{subequations}

\subsubsection{Projection of the Poisson bracket}
Assume now that when functionals dependent only on the less detailed variables $\yy$ are plugged into the Poisson bracket, the resulting expression depends only on variables $\yy$ and it is referred to as the less-detailed Poisson bracket $^\downarrow\{\bullet,\bullet\}$, i.e.
\begin{equation}
\{A(\yy),B(\yy)\} = A_{y^a} \underbrace{\pi^a_i L^{ij} \pi^b_j}_{=\dL^{ab}(\yy)} B_{y^b} = ^\downarrow\{A,B\}.
\end{equation}
That is the case for example when $\yy$ are the hydrodynamic fields of density, momentum density and entropy density and $\xx$ the one-particle distribution functions. Poisson brackets can be often derived from more detailed Poisson brackets by such projections, see e.g. \cite{PhysicaD-2015}.

The projections are naturally made in the so called energetic representation, see \cite{Callen}, where for example in the case of hydrodynamic field entropy density is among the state variables instead of energy density. On the other hand, in the entropic representation the field of energy density is among the state variables. In that case, it is natural to employ the MaxEnt principle to find the least biased estimate of the detailed variable $\xx$ based on the knowledge of the less detailed variables $\yy$. In the case of the energetic representation, the MaxEnt principle is replaced by the principle of minimal energy: find $\xx$ such that energy $E(\xx)$ is minimal subject to the knowledge of field $\yy$ that include the less detailed entropy density. This can be expressed by Legendre transformation
\begin{subequations}
\begin{equation}\label{eq.dE}
\frac{\partial}{\partial \xx}\left(-E(\xx) + \yy^*\cdot \yy(\xx)\right)\Big|_{\tilde{\xx}(\yy)} = 0,
\end{equation}
which gives the dependence $\tilde{\xx}(\yy^*)$ and a new functional 
\begin{equation}
\dE^*(\yy^*) = -E(\tilde{\xx}(\yy)) + \yy^*\cdot \yy(\tilde{\xx}(\yy^*)),
\end{equation}
the conjugate lower energy, which is a functional of conjugate less detailed (lower) variables $\yy^*$. The energy on the lower level is then obtained by a subsequent Legendre transformation 
\begin{equation}
\frac{\partial}{\partial \yy^*}\left(-\dE^*(\yy^*) + \yy^*\cdot \yy\right)\Big|_{\yy^*(\yy)} = 0
\end{equation}
and
\begin{equation}
\dE(\yy) = -\dE^*(\yy^*(\yy)) + \yy^*(\yy)\cdot \yy.
\end{equation}
These equations imply, in particular, that 
\begin{equation}
\dE^*_{y^*_a} = y^a \mbox{ and } \dE_{y^a} = y^*_a.
\end{equation}
From Eq. \eqref{eq.dE} it then follows that 
\begin{equation}\label{eq.dE.ddE}
E_{x^i}\Big|_{\tilde{\xx}(\yy)} = y^*_a \frac{\partial \pi^a}{\partial x^i} =  \frac{\partial \dE}{\partial y^a} \frac{\partial \pi^a}{\partial x^i},
\end{equation}
which will be useful later.
\end{subequations}

Assuming that we have already projected the Poisson bracket to the lower level of description, evolution of the less detailed state variables is given by  
\begin{equation}\label{Ham:eq.y.evo}
\dot{y}^a = \underbrace{ \dL^{ab} \dE_{y^b}}_{=\dJ^a}\qquad \mbox{or}\qquad \dot{A}=^\downarrow\{A,\dE\} \forall A(\yy).
\end{equation}
The Taylor expansion in time on the less-detailed level of description then yields
\begin{equation}\label{Ham:eq.y.Taylor}
A_a y^a(t + \tau) = A_a y^a(t) + \tau A_a \dL^{ab} \dE_{y^b} + \frac{\tau^2}{2} A_a\frac{\partial}{\partial y^c}\left(\dL^{ab}\dE_{y^b}\right)\dL^{cd}\dE_{y^d}  + \OO(\tau^3) \quad \forall A_a.
\end{equation}
This last equation is the analogy of Eq. \eqref{eq.LowerLevel_expansion}, and can be regarded as evolution equation
\begin{equation}\label{eq.y.reg}
\frac{d y^a}{d \tau} =\dL^{ab} \dE_{y^b} + \frac{\tau}{2} \frac{\partial}{\partial y^c}\left(\dL^{ab}\dE_{y^b}\right)\dL^{cd}\dE_{y^d},
\end{equation}
where the right hand side consists of a reversible (first) and an irreversible (second) term. This is again a \textbf{self-regularization} of evolution \eqref{Ham:eq.y.evo}.

\subsubsection{Comparing the solutions on different levels}
The next step in the Ehrenfest reduction is to alter the right hand side of the less-detailed evolution equation so that the solution is closer to the solution on the detailed level of description. 

\begin{figure}[ht!]
\begin{center}
\includegraphics[scale=0.3]{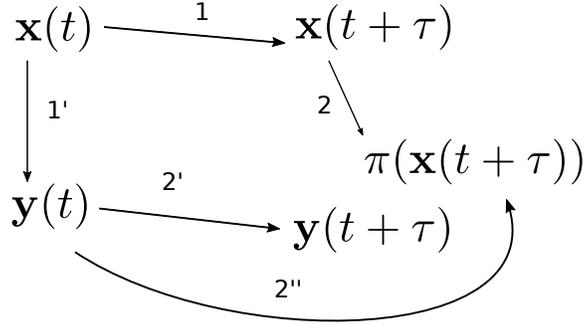}
\caption{\label{fig.HamEhr}Hamiltonian interpretation of the Ehrenfest reduction. Step 1: The exact more detailed evolution equations are first solved formally to obtain their solution at time $t+\tau$. Step 2: This solution, $\xx(t+\tau)$, is then projected to the less detailed level to obtain $\proj(\xx(t+\tau))$. Step 1': Alternative route is to first project $\xx(t)$ to $\yy(t)$. Step 2': The less detailed evolution equation (generated by the projection of the Poisson bracket) then takes $\yy(t)$ and gives $\yy(t+\tau)$. Step 2'': We have thus $\proj(\xx(t+\tau))$ and $\yy(t+\tau)$, which should ideally be equal, but they are typically not. The value $\proj(\xx(t+\tau))$ is of course more precise because it is constructed from the detailed evolution equations. To make the value $\yy(t+\tau)$ more precise, the less detailed evolution equations are altered by adding the difference between the self-regularized detailed and less-detailed equations. Such equations for $\yy$ are then the reduced Ehrenfest equations.}
\end{center}
\end{figure}

More precisely, Taylor expansion \eqref{Ham:eq.x.Taylor} gives an approximation of $x^i(t+\tau)$ provided $x^i(t)$ is known. Similarly, Taylor expansion \eqref{Ham:eq.y.Taylor} gives an approximation of $y^a(t+\tau)$ provided $y^a(t)$ is known. The value $y^a(t+\tau)$ can be, however, approached by two ways: (i) Projection from $\xx(t)$ to $\yy(t)$ and subsequent evolution by Taylor expansion \eqref{Ham:eq.y.Taylor} or (ii) evolution by expansion \eqref{Ham:eq.x.Taylor} and subsequent projection from $\xx(t+\tau)$ to $\yy(t+\tau)$. The second route is of course more precise, but one has to solve the detailed evolution equations for $\xx$. Therefore, the first route is more suitable, but it needs a further correction to make it closer to the second route. An additional term is thus added to expansion \eqref{Ham:eq.y.Taylor}, which makes the two routes equivalent up to the given order of $\tau$. This procedure is summarized in Fig. \ref{fig.HamEhr}.

Let us thus compare the self-regularized evolutions \eqref{Ham:eq.x.Taylor} and \eqref{Ham:eq.y.Taylor}. The first terms of the expansions  are already related by the projection because (using Eq. \eqref{eq.dE.ddE})
\begin{equation}
\pi^a_i L^{ij} E_{x^j}|_{\tilde{\xx}(\yy)} = 
\pi^a_i L^{ij} \pi^b_j \dE_{y^b} = \dL^{ab}\dE_{y^b}.
\end{equation}

The second terms of the expansions are no longer related by such a projection. Therefore, one has to add the difference between these two second terms. The regularized evolution equation for $\yy$ then becomes
\begin{equation}\label{Ham:eq.y.final}
\dot{y}^a = \dL^{ab}E_{y^b} + \frac{\tau}{2}\left(
\pi^a_i \frac{\partial}{\partial x^k}\left(L^{ij}E_{x^j}\right)L^{kl}E_{x^l} 
-\frac{\partial}{\partial y^c}\left(\dL^{ab}\dE_{y^b}\right)\dL^{cd}\dE_{y^d}\right)\Big|_{\tilde{\xx}(\yy)},
\end{equation}
which is the less-detailed evolution equation obtained by the Ehrenfest reduction, analogy of Eq. \eqref{eq.evo.y.final}.

\subsubsection{The special case of constant Poisson bivector}
Hamilton canonical equations of classical mechanics are generated by the canonical Poisson bracket or canonical Poisson bivector, 
\begin{equation}\label{eq.PB.can}
\{A,B\}^{(can)} = A_\rr \cdot B_\pp - A_\pp \cdot B_\rr, 
\qquad \LL =
\begin{pmatrix}
0 & 1\\
-1 & 0
\end{pmatrix}.
\end{equation}
The canonical Poisson bivector is a constant matrix. 

The Poisson bracket for hydrodynamic density and momentum density $(\rho,\uu)$ is not a canonical one. However, it can be casted into the canonical form by transformation from $(\rho,\uu)$ into the so called Clebsch variables, see \cite{Clebsch} and \cite{Marsden1983}, or even $(\rho,\uu,s)$ as in \cite{Bedeaux-Mazur}. Constant Poisson bivectors can be thus often met.

If the bivector is constant, expansions of the solutions \eqref{Ham:eq.x.Taylor} and \eqref{Ham:eq.y.Taylor} obtain a particularly simple form, 
\begin{subequations}
\begin{align}
 x^i(t + \tau) &= x^i(t) + \tau  L^{ij} E_{x^i} + \frac{\tau^2}{2} \underbrace{L^{ij}\frac{\delta^2 E}{\delta x^j x^k}L^{kl}}_{=-M^{ik}}E_{x^l}  + \OO(\tau^3)\\
 y^a(t + \tau) &= y^a(t) + \tau \dL^{ab}E_{y^b} + \frac{\tau^2}{2} \underbrace{\dL^{ab}\frac{\delta^2 E}{\delta y^c y^b}\dL^{cd}}_{=-\dM^{ad}}E_{y^d}  + \OO(\tau^3),
\end{align}
\end{subequations}
which gives evolution equations
\begin{subequations}\label{Ham:eq.can.Taylor}
\begin{align}
 \dot{x}^i(t) &= L^{ij} E_{x^i} - \frac{\tau}{2} M^{ij}E_{x^j}\\
 \dot{y}^a(t) &= \dL^{ab} E_{y^b} - \frac{\tau}{2} \dM^{ab}E_{y^b}.
\end{align}
\end{subequations}
Matrices $\MM$ and $\dMM$ are clearly symmetric. Moreover, it follows from convexity of energy, second differential of which is thus a positive definite matrix, that
\begin{equation}
v_i M^{il} v_l= -v_i L^{ij} E_{x^j x^k} L^{kl} v_l = L^{ji}v_i E_{x^j x^k} L^{kl} v_l \geq 0 \quad\forall \mathbf{v},
\end{equation}
and the matrices $\MM$ and $\dMM$ are thus symmetric positive semidefinite. In particular, it means that energy is destroyed in both equations \eqref{Ham:eq.can.Taylor}. Evolution equations \eqref{Ham:eq.can.Taylor} can be interpreted as self-regularizations of the original completely reversible evolution equations \eqref{Ham:eq.x.evo} and \eqref{Ham:eq.y.evo}.

The Ehrenfest reduced evolution equation is then 
\begin{equation}\label{Ham:eq.y.final.can}
 \dot{y}^a(t) = \dL^{ab} E_{y^b} - \frac{\tau}{2} \left(\pi^a_i M^{ij}|_{\tilde{\xx}(\yy)}E_{x^j}|_{\tilde{\xx}(\yy)} - \dM^{ab}E_{y^b}\right),
\end{equation}
which is the analogy of Eq. \eqref{eq.evo.y.final}.

\subsubsection{Canonical Hamiltonian system}
For instance the canonical Poisson bivector \eqref{eq.PB.can} is constant. The canonical Poisson bracket, that provides natural kinematics on cotangent bundles, is formed from the Poisson bivector. 

Let the variables on a cotangent bundle be denoted by $(\rr,\pp)$, which can be interpreted as position and momentum of a particle. Evolution of these variables is then given by
\begin{equation}
\begin{pmatrix}
\dot{\rr}\\
\dot{\pp}
\end{pmatrix}
=
\begin{pmatrix}
0 & 1\\
-1 & 0
\end{pmatrix}
\cdot
\begin{pmatrix}
H_\rr \\
H_\pp
\end{pmatrix}
\end{equation}
with $H(\rr,\pp)$ being energy of the particle in a static external field, e.g.
\begin{equation}\label{eq.H.particle}
H(\rr,\pp) = \frac{\pp^2}{2m} + V(\rr).
\end{equation}

The self-regularized evolution of $(\rr,\pp)$ is then, according to Eq. \eqref{Ham:eq.can.Taylor}, 
\begin{equation}
\begin{pmatrix}
\dot{\rr}\\
\dot{\pp}
\end{pmatrix}
=
\begin{pmatrix}
0 & 1\\
-1 & 0
\end{pmatrix}
\cdot
\begin{pmatrix}
H_\rr \\
H_\pp
\end{pmatrix}
-\frac{\tau}{2}
\underbrace{
\begin{pmatrix}
H_{\pp\pp} & -H_{\pp\rr}\\
-H_{\rr\pp} & H_{\rr\rr}
\end{pmatrix}
}_{=\MM}
\cdot
\begin{pmatrix}
H_\rr \\
H_\pp
\end{pmatrix}
.
\end{equation}
For the particular choice of Hamiltonian \eqref{eq.H.particle}, the evolution equations become
\begin{equation}
\begin{pmatrix}
\dot{\rr}\\
\dot{\pp}
\end{pmatrix}
=
\begin{pmatrix}
\frac{\pp}{m} \\
-V_\rr
\end{pmatrix}
-\frac{\tau}{2}
\underbrace{
\begin{pmatrix}
\frac{1}{m} & 0\\
0 & V_{\rr\rr}
\end{pmatrix}
}_{=\MM}
\cdot
\begin{pmatrix}
V_\rr \\
\frac{\pp}{m}
\end{pmatrix}
=
\begin{pmatrix}
\frac{\pp}{m} \\
-V_\rr
\end{pmatrix}
-\frac{\tau}{2m}
\begin{pmatrix}
V_\rr\\
V_{\rr\rr}\cdot\pp
\end{pmatrix}
\end{equation}
This is the self-regularized evolution of the particle.

Assuming that the external field $V(\rr)$ is convex and that it has a minimum, the Hamiltonian (or energy) is also convex. Consequently, the dissipative matrix $\MM$ is positive-semidefinite, and energy is dissipated in course of the evolution as
\begin{equation}
\dot{H} = -\frac{\tau}{2}\left(\frac{1}{m}(V_\rr)^2 + \frac{\pp}{m}\cdot V_{\rr\rr} \cdot \frac{\pp}{m}\right)\leq 0,
\end{equation}
and the evolution stops at the position and momentum satisfying
\begin{equation}
V_\rr = 0 \qquad\mbox{and}\qquad \pp = 0, 
\end{equation}
which is the equilibrium position of the particle within the field as well as the position of lowest total energy.

Let us now consider a less detailed level of description where only the position $\rr$ plays the role of state variable. The mapping from $(\rr,\pp)$ to $(\rr)$ is the projection between the two levels of description. The Poisson bivector on the less detailed description is a 1x1 matrix, i.e. a number, and it is given by 
\begin{equation}
(1,0) \cdot \LL \cdot
\begin{pmatrix}
1\\
0
\end{pmatrix} = 0.
\end{equation}
The less detailed Poisson bivector is identically zero. The less detailed evolution is thus zero as well as the self-regularized less detailed evolution. 

But the less detailed evolution derived by the Ehrenfest reduction, Eq. \eqref{Ham:eq.y.final.can}, is not zero and reads
\begin{equation}
\dot{\rr} = -\frac{\tau}{2m}(1,0)\cdot 
\begin{pmatrix}
V_\rr\\
V_{\rr\rr}\cdot\tilde{\pp}(\rr)
\end{pmatrix}
=-\frac{\tau}{2m} V_\rr.
\end{equation}
This reduced evolution tends again to the minimum of the external field.


\section{From Vlasov to mechanical equilibrium}
Now we are coming to the main point of the paper - the reduction of the Vlasov equation to mechanical equilibrium. The reduced equations will naturally exhibit the tendency to spatially homogeneous equilibrium, which can be interpreted as a manifestation of Landau damping. But let us first define the projection from the Vlasov level of description to the level of mechanical equilibrium.

\subsection{Projection}
Since the particle density approaches the equilibrium in a strong sense, we can expect that if entropy were a function of the density, it could approach its maximum. Therefore, it is sensible to regard Vlasov equation from a less detailed (or lower) level of description, for example the level of mechanical equilibrium where state variables are
\begin{subequations}\label{eq.proj}
\begin{eqnarray}
\mbox{particle density: }&&\rho(\rr) = \int d\pp  m f(\rr,\pp)\\
\mbox{ kinetic energy density: }&&\eps(\rr) = \int d\pp \frac{\pp^2}{2m}f(\rr,\pp).
\end{eqnarray}
\end{subequations}
The level of description where $\rho$ and $\eps$ play the role of state variables is referred to as the level of mechanical equilibrium (motivated by \cite{dGM}) because momentum is not present among the state variables.

Let us justify the reduction to the level of mechanical equilibrium from the perspective of rigorous mathematical results summarized in Sec. \ref{sec.introduction}. Consider a particular choice of the test function that is a tensor product $\varphi(\rr,\pp)=\pp\otimes\tilde{\varphi}(\rr)$, where $\tilde{\varphi}$ is smooth. Then we have that 
\begin{equation*}
 \int h(t,\rr,\pp) \varphi(\rr,\pp) d\rr d\pp=\int d\rr \tilde{\varphi}(\rr)\int d\pp~\pp h(t,\rr,\pp)\rightarrow \int d\rr \tilde{\varphi}(\rr) \int d\rr' d\pp~\pp h_i(\rr',\pp).
\end{equation*}
As this relation holds for all smooth $\tilde{\varphi}$ and particularly for smooth cutoff functions (convolution of a characteristic function with a mollifier) we can observe that
\begin{equation*}
\lim_{t\rightarrow +\infty} \int dv~v h(t,x,v) = \int dy dv~v h_i(y,v),
\end{equation*}
which is a constant in space. Therefore not only the macroscopic density but also the macroscopic momentum,
\begin{equation}
\uu(t,\rr) = \int d\pp \pp f(t,\rr,\pp),
\end{equation}
tend, as $t\rightarrow\infty$, to a spatially homogeneous state.  Hence it is natural to consider the mechanical equilibrium as a reasonable level of description for large time behavior of the system. We shall explore this Landau damping via pertinent Ehrenfest reduction that is able to reveal irreversible behavior on the reduced level of description.

Total energy \eqref{eq.energy} can be expressed as
\begin{equation}
E = \int d\rr \eps(\rr) + \frac{1}{2} \int d\rr \int d\rr' \frac{\rho(\rr)}{m} V(|\rr-\rr'|) \frac{\rho(\rr')}{m}.
\end{equation}
Mass and energy density are thus obtained from the distribution function by projection \eqref{eq.proj}.

On the other hand, knowledge of fields $\rho$ and $\eps$ is insufficient for exact reconstruction of the distribution function. What is the estimate of the distribution function when only those two fields are accessible? The answer is the maximum entropy principle (MaxEnt), which tells that the least biased estimate of the distribution function is the solution to equation
\begin{equation}
\frac{\delta}{\delta f} \left(-S^{(B)}(f) + \int d\rr \rho^*(\rr) \rho(f) + \int d\rr \eps^*(\rr) \eps(f)\right)= 0,
\end{equation}
where fields $\rho^*$ and $\eps^*$ play the role of Lagrange multipliers ensuring constraints \eqref{eq.proj}, but can be also given geometrical interpretation \cite{RedExt}. Solving this equation leads to 
\begin{equation}
\tilde{f}(\rho,\eps;\pp) = \frac{1}{h^3}\exp\left(-\frac{m\rho^*(\rho,\eps)}{k_B}\right)\exp\left(-\frac{\eps^*(\rho,\eps)}{k_B} \frac{\pp^2}{2m}\right),
\end{equation}
and by using the constraints \eqref{eq.proj}, the Lagrange multipliers can be expressed as functions of $\rho$ and $\eps$, which finally leads to
\begin{equation} \label{eq.f.quasieq}
\tilde{f}(\rho,\eps;\pp) = \frac{\rho}{ m^4} \left(\frac{3}{4\pi} \frac{\rho}{\eps}\right)^{3/2} \exp\left(-\frac{3}{2}\frac{\rho}{\eps} \frac{\pp^2}{2m}\right),
\end{equation}
which is the MaxEnt estimate of the distribution function subject to the knowledge of field $\rho$ and $\eps$.

Boltzmann entropy evaluated at the MaxEnt distribution function then becomes the entropy on the level of mechanical equilibrium, 
\begin{equation}\label{eq.S.ME}
S^{(ME)}(\rho,\eps) = -k_B \int d\rr \frac{\rho}{m}\left(-\frac{5}{2} + \frac{3}{2}\ln\left(\frac{3h^2}{4\pi m^2}\frac{\rho}{\eps}\right) + \ln \frac{\rho}{m}\right),
\end{equation}
which is in fact the local equilibrium version of Sackur-Tetrode equation \cite{Callen}.

Having constructed the projection from $f$ onto the level of mechanical equilibrium, a mapping from mechanical equilibrium back into the space of distribution functions and the implied mechanical-equilibrium entropy, we have specified the static relations between the two levels of description (Vlasov and mechanical equilibrium). Let us now have a look at dynamical relations between the levels, i.e. how evolution on one level corresponds with evolution on the another level.

\subsection{Construction of the reduced evolution}
Let us now apply the Ehrenfest reduction to the passage from Vlasov level of description onto the level of mechanical equilibrium. State variables are identified as
\begin{equation}
\xx = f(\rr,\pp) \qquad\mbox{and} \qquad \yy = (\rho(\rr),\eps(\rr)).
\end{equation}
The more detailed evolution equation is the Vlasov equation \eqref{eq.Vlasov}. The sought reduced evolution equations are evolution equations for $\rho$ and $\eps$.

The zeroth approximation is zero, which readily follows from the observation that 
\begin{equation*}
	J(\tilde{x}(y)) =  - \frac{\pp}{m} \cdot \frac{\partial \tilde f}{\partial \rr} - \FF(\tilde f)(t,\rr)\cdot \frac{\partial \tilde f}{\partial \pp},
\end{equation*}
where $\tilde f(\rho,\eps;\pp)$ is the quasi-equilibrium state calculated in \eqref{eq.f.quasieq}. The zeroth approximation of the macroscopic density then follows from the corresponding projection of the last expression for $J(\tilde{x}(y))$ which is
\begin{equation*}
\rho^{(0)}=\langle \pi_\rho, J(\tilde{x}(y)) \rangle = \int d\pp m \left[ - \frac{\pp}{m} \cdot \frac{\partial \tilde f}{\partial \rr} - \FF(\tilde f)(t,\rr)\cdot \frac{\partial \tilde f}{\partial \pp} \right].
\end{equation*}
The first term vanishes as it is an odd function in all $p_k$ and the second term can be simply integrated to zero as $\FF(t,\rr)$ is independent on momentum $\pp$. Analogously one can show that the projection yielding energy density $\eps$ vanishes as well on $J(\tilde{x}(y))$ and hence the zeroth approximation $\eps^{(0)}$ is zero.

The first approximation that corresponds to irreversible evolution can again be calculated from the relations \eqref{eq.R1} identified above. As the zeroth approximation is zero, i.e. $\tilde{\JJ}=0$, only the first terms contribute to the first order correction. Let us start by calculating the variation of $\delta J(\xx)/\delta \xx$
\begin{align*}
	J(f(\rr'',\pp'')&=-\frac{p_k''}{m}\frac{\partial f}{\partial r_k''}(\rr'',\pp'')+\frac{\partial f}{\partial p_k''}(\rr'',\pp'') \int d\rr'\int d\pp' \frac{\partial V(|\rr''-\rr'|)}{\partial r_k''} f(\rr',\pp')=\\
	&=\int d\rr \int d\pp ~ \delta(\rr-\rr'')\delta(\pp-\pp'')\cdot\nonumber\\
        &\qquad\cdot\left[ -\frac{p_k}{m}\frac{\partial f(\rr,\pp)}{\partial r_k} + \frac{\partial f(\rr,\pp)}{\partial p_k} \int d\rr'\int d\pp' \frac{\partial V(|\rr-\rr'|)}{\partial r_k} f(\rr',\pp') \right]
\end{align*}
where $\frac{\partial f(\rr'',\pp'')}{\partial r_k''}\Big|_{\rr''=\rr,\pp''=\pp}=\frac{\partial f(\rr,\pp)}{\partial r_k}$ was used. Hence the variation of $\delta J(\xx)/\delta \xx$ is
\begin{multline*}
 \frac{\delta J(f(\rr'',\pp''))}{\delta f(\rr,\pp)} = \partial_{r_k}(\delta(\rr-\rr'')) \delta(\pp-\pp'') \frac{p_k}{m} - \delta(\rr-\rr'') \left(\partial_{p_k} \delta(\pp-\pp'')\right)\cdot\nonumber\\
 \cdot\int d\rr' \int d\pp' \frac{\partial V (|\rr-\rr'|)}{\partial r_k} f(\rr',\pp')+\\
 +\underbrace{\int d\rr'\int d\pp' \delta(\rr'-\rr'') \delta(\pp'-\pp'') \frac{\partial f(\rr',\pp')}{\partial p_k'} \frac{\partial V(|\rr'-\rr|)}{\partial r_k'}}_{=\frac{\partial f(\rr'',\pp'')}{\partial p_k''} \frac{\partial V(|\rr''-\rr|)}{\partial r_k''}}.
\end{multline*}

The contribution to the macroscopic evolution equations is the projection of
\begin{align*}
\langle\frac{\delta J(\xx(t))}{\delta \xx}&\Big|_{\xx=\tilde{\xx}(y)}, J(\xx(t))\Big|_{\xx=\tilde{\xx}(y)}\rangle = \int d\rr \int d\pp \frac{\delta J(f(\rr'',\pp'')}{\delta f(\rr,\pp)} J(f(\rr,\pp))\Big|_{f=\tilde f}=\\
&=\int d\rr \int d\pp \frac{p_k''}{m}\frac{p_j''}{m} \frac{\partial^2 \tilde f(\rr'',\pp'')}{\partial r_k''\partial r_j''}+\frac{p_k''}{m}  \frac{\partial^2 \tilde f(\rr'',\pp'')}{\partial r_k''\partial p_j''} F_j(\rr'')\nonumber\\
&-\frac{p_k''}{m} \frac{\partial \tilde f(\rr'',\pp'')}{\partial p_j''}\int d\rr'''d\pp'''\frac{\partial^2 V(\rr''-\rr'''|)}{\partial r_k''\partial r_j''} \tilde f(\rr''',\pp''')+\\
&+\frac{\delta_{jk}}{m}\frac{\partial \tilde f(\rr'',\pp'')}{\partial r_j''}F_k(\rr'') + \frac{p_j''}{m} \frac{\partial^2 \tilde f(\rr'',\pp'')}{\partial p_k''\partial r_j''} F_k(\rr'') + \frac{\partial^2 \tilde f(\rr'',\pp'')}{\partial p_k''\partial p_j''} F_k(\rr'') F_j(\rr'')+\\
&+\frac{\partial \tilde f(\rr'',\pp'')}{\partial p_k''} \frac{p_j}{m} \tilde f(\rr,\pp) \frac{\partial^2 V(|\rr''-\rr|)}{\partial r_j \partial r_k''}.
\end{align*}

The first correction to the macroscopic evolution equation for density $\frac{\partial \rho}{\partial t}(t,\rr'')$ follows from projection $\pi_\rho=\int d\pp''  m$ applied to this last relation and the particular form of quasi-equilibrium $\tilde f$ from \eqref{eq.f.quasieq} is employed
\begin{align*}
	\frac{\partial \rho}{\partial t}(t,\rr'') &=  \tau m \int d\pp'' \left(\frac{p_k''}{m}\frac{p_j''}{m} \frac{\partial^2 \tilde f(\rr'',\pp'')}{\partial r_k''\partial r_j''}\right.\nonumber\\
        &\left.-\frac{\delta_{jk}}{m} \frac{\partial \tilde f}{r_k''} F_j +\frac{\delta_{jk}}{m}\tilde f(\rr'',\pp'') \int d\rr'''d\pp'''\frac{\partial^2 V(\rr''-\rr'''|)}{\partial r_k''\partial r_j''} \tilde f(\rr''',\pp''')\right)=\\
	&= \tau \left[2\frac{\partial^2}{\partial r_k''\partial r_j''}\underbrace{\int d\pp'' \frac{p_k'' p_j''}{2m} \tilde f(\rr'',\pp'')}_{=0\text{ for }k\neq j}-F_j\frac{\delta_{jk}}{m}\frac{\partial \rho(\rr'')}{\partial r_k''}- \frac{\delta_{jk}}{m}\rho(\rr'')\frac{\partial F_j(\rr'')}{\partial r_k''}\right]
\end{align*}
where the integral in the last equation vanishes unless $k=j$ due to symmetry of $\tilde f$ (an even function in $p_i$). Its nonzero value is  $\eps/3$ as can be seen from the projections.

In exactly the same manner one can proceed in obtaining the first (irreversible) approximation of the macroscopic evolution equation for energy density. Finally, the reduced evolution equations become
\begin{subequations} \label{eq.Macro.Final}
\begin{eqnarray}
\frac{\partial \rho}{\partial t} &=& \frac{\tau}{2} \left[\frac 2 3  \Delta \eps - \nabla\cdot(\rho  \FF^{(ME)})\right],\\
\frac{\partial \eps}{\partial t} &=& \frac{\tau}{2} \left[\frac{10}{9}   \Delta\left(\rho \left(\frac{\eps}{\rho}\right)^2\right)  - \nabla\cdot( \eps  \FF^{(ME)}) +  \rho \left( \FF^{(ME)}\right)^2	\right].
\end{eqnarray}
\end{subequations}
where the macroscopic force $\FF^{(ME)}$ is related to the macroscopic variables via
\begin{eqnarray}
  F^{(ME)}_j(\rr'') &=&  \frac 1 m F_j|_{f=\tilde f} = - \frac 1 m\int d\rr' \int d\pp' \frac{\partial V(|\rr''-\rr'|)}{\partial r_j''} \tilde{f}(\rr',\pp')\nonumber\\
 &=& - \frac{1}{m^2}\int d\rr' \frac{\partial V(|\rr''-\rr'|)}{\partial r_j''} \rho(\rr').
\end{eqnarray}
In summary, the reduced evolution on the level of mechanical equilibrium is irreversible although the detailed evolution (Vlasov equation \eqref{eq.Vlasov}) was completely reversible, and entropy on the level of mechanical equilibrium is expected to grow in time although the Boltzmann entropy is conserved by the Vlasov equation.

\subsection{Features of the reduced evolution}

\subsubsection{Dissipativity of reduced evolution}
To assess the growth of the reduced entropy, we evaluate the following expression
\begin{equation}
\dot{S}^{(ME)} = \int d\rr S_\rho(\rr) \frac{\partial \rho}{\partial t} + S_\eps(\rr) \frac{\partial \eps}{\partial t}, 
\end{equation}
and obtain
\begin{equation}\label{eq.entprod}
\dot{S}^{(ME)} = \frac{\tau k_B}{2   m} \int d \rr \left[ \frac{5}{3} \eps \rho^{-2} (\partial_k \rho)^2+\frac{7}{3}\eps^{-1}(\partial_k \eps)^2 -\frac{10}{3}\rho^{-1} \partial_k\eps\partial_k\rho- \FF^{ME}.\nabla \rho +\frac{3}{2}  \eps^{-1}\rho^2\left( \FF^{ME}\right)^2\right].
\end{equation}

The third and fourth terms in formula \eqref{eq.entprod} do not have any definite sign in contrast to the other terms, which are all non-negative. The third term can be estimated from above by the H\"{o}lder and Young inequalities as
\begin{eqnarray}
\Big|\int\diff\rr -\frac{10}{3}\frac{1}{\rho}\nabla\rho\cdot\nabla\eps\Big|&=& 
\frac{10}{3}\Big|\int\diff\rr \left(\frac{2}{\sqrt{5}}\frac{\sqrt{\eps}}{\rho}\nabla\rho\right)\cdot\left(\frac{\sqrt{5}}{2}\frac{\nabla\eps}{\sqrt{\eps}}\right)\Big| \nonumber\\
&\leq& \frac{5}{3}\left(\int\diff\rr \frac{4}{5}\frac{\eps}{\rho^2}(\nabla\rho)^2 + \int\diff\rr \frac{5}{4}\frac{(\nabla\eps)^2}{\eps}\right)
\end{eqnarray}
while the fourth term as
\begin{eqnarray}
\Big|\int\diff\rr \FF^{(ME)}\cdot\nabla\rho\Big| &=& \Big|\int\diff\rr \left(\frac{\sqrt{3}}{\sqrt{2}}\frac{\FF^{(ME)}\rho}{\sqrt{\eps}}\right)
\cdot\left(\frac{\sqrt{2}}{\sqrt{3}}\frac{\sqrt{\eps}\nabla\rho}{\rho}\right)\nonumber\\
&\leq& \sqrt{\int\diff\rr \frac{3}{2}\frac{\left(\FF^{(ME)}\right)^2 \rho^2}{\eps}}
\cdot \sqrt{\int\diff\rr \frac{2}{3} \frac{(\nabla\rho)^2 \eps}{\rho^2}}\nonumber\\
&\leq& \int\diff\rr \frac{3}{4} \frac{\left(\FF^{(ME)}\right)^2 \rho^2}{\eps}
+ \int\diff\rr \frac{1}{3} \frac{\left(\nabla\rho\right)^2 \eps}{\rho^2}.
\end{eqnarray}
Plugging these two estimates into the formula for entropy production \eqref{eq.entprod} leads to the inequality
\begin{equation}
\dot{S}^{(ME)} \geq \frac{\tau k_B}{2 m} \int \diff \rr \frac{3}{4}\frac{\left(\FF^{(ME)}\right)^2\rho^2}{\eps}+\frac{1}{4}\frac{(\nabla\eps)^2}{\eps}\geq 0,
\end{equation}
which means that the reduced entropy production is always non-negative. The reduced evolution equations are thus compatible with the second law of thermodynamics.

\subsubsection{Homogeneous equilibrium solution}

A natural question now arises whether there exists a (spatially) homogeneous solution of the macroscopic evolution equations \eqref{eq.Macro.Final} describing the system in mechanical equilibrium, i.e. a solution to 
\begin{align*}
0&=\nabla. \FF^{ME},\\
0&=-\eps \nabla. \FF^{ME}+ \left( \FF^{ME}\right)^2 \rho.
\end{align*}
Hence one immediately arrives at a necessary condition for the existence of a homogeneous solution which states that $ \FF^{ME}=0$, perfectly corresponding to the observation of Villani et al that $ \FF^{ME}$ strongly tends to zero with large times (as perturbations decay). With the knowledge of the particular form of the quasi-equilibrium $\tilde f$ and with the assumption of homogeneous $\rho,~\eps$ we may proceed further
\begin{equation*}
 F^{ME}_k=0=\frac{\partial}{\partial r_k} \int d\rr' \int d\pp' V(|\rr-\rr'|) \tilde f(\rho,\eps,\pp)=const \frac{\partial}{\partial r_k} \int d\rr' V(|\rr-\rr'|)
\end{equation*}
to observe that a homogeneous solution exists iff 
\begin{align*}
\Omega=\R^3 \mbox{ and } &\int_\Omega d\rr V(|\rr|)<+\infty,\\
\Omega \mbox{ is periodic as a system } & \mbox{(e.g. a torus) } \Rightarrow \frac{\partial}{\partial r_k} \int d\rr' V(|\rr-\rr'|)=0
\end{align*}
revealing a nontrivial condition for the interacting potential  for unbounded regions (again cf. findings about linear and nonlinear Landau damping). 

In such cases, any constant is a solution. The choice of this homogeneous solution is not, however, arbitrary. It follows from the initial condition and the evolution equation accompanied by entropy functional guiding this process.
To prove mathematically that the reduced evolution equations tend to this homogeneous solution is beyond the scope of this paper (if can be done at all: strongly coupled nonlinear parabolic PDEs). However, the approach of $\rho$, precise evolution of which is given by the Vlasov equation, to a homogeneous equilibrium was proven by Villani and Mouhot as a manifestation of nonlinear Landau damping. The approach of $\rho$ to a homogeneous equilibrium by means of the reduced evolution equations \eqref{eq.Macro.Final} can be anticipated. 

Is the homogeneous solution the state where maximum of the mechanical equilibrium entropy is reached? The system of evolution equations is accompanied by entropy functional \eqref{eq.S.ME} accompanying the evolution of mechanical equilibrium variables $\rho$ and $\eps$ towards the equilibrium values. Maximization of the reduced entropy with respect to constraints
\begin{equation}
 M(\rho,\eps)=\int d\rr \rho(\rr),~E(\rho,\eps)=\int d \rr \eps+\frac{1}{m^2}\frac 1 2 \int d\rr \int d\rr' \rho(\rr) V(|\rr-\rr'|) \rho(\rr')
\end{equation}
can be written as
\begin{subequations}
\begin{align}
&\frac{\delta}{\delta \rho}\left(-S^{(ME)} + E^* \cdot E(\rho,\eps) + M^*\cdot  M(\rho,\eps)\right) = 0\\
&\frac{\delta}{\delta \eps}\left(-S^{(ME)} + E^* \cdot E(\rho,\eps) + M^*\cdot  M(\rho,\eps)\right) = 0.
\end{align}
\end{subequations}
These equations lead to 
\begin{subequations} \label{eq.ConjugEq}
\begin{eqnarray}
 E^*&=&\frac 3 2 k_B \frac{\rho}{ m\eps},\\
 \label{eq.Nstar}M^*&=& -k_B \frac{3}{2 m}\ln\frac{3 h^2}{4 \pi  m^2}\frac{\rho}{\eps} - \frac{k_B}{m} \ln \frac{\rho}{ m} -E^*\frac{1}{m^2} \int d\rr' V(|\rr-\rr'|) \rho(\rr').
\end{eqnarray}
\end{subequations}
Therefore, there is a relation between particle density and energy density (in equilibrium) via a constant $\alpha$
\begin{equation*}
	\eps=\alpha  \rho
\end{equation*}
and consequently 
total energy becomes for homogeneous distribution of particles
\begin{equation*}
E=\alpha M+\frac{1}{2m^2} \nu \barn M,
\end{equation*}
where for $\nu=\int d\rr V(|\rr|)$ and $\barn = M/|\Omega|$, $\Omega$ being the spatial integration domain.  

To proceed further and comply with the definitions of energies above, one can see that we need to restrict ourselves to finite domains. Hence we are forced further to assume that $\Omega$ is a finite domain, e.g. a torus, in line with the assumptions in the work of Mouhot and Villani \cite{Mouhot2011}.
 This identifies constant $\alpha$ as function of $E$, $M$, $\Omega$ and of the potential $V$.  Moreover, Eq. \eqref{eq.Nstar} gives the Lagrange multiplier $M^*$ as a function of $E$, $M$ and $\Omega$ as 
\begin{equation}
 M^* = -k_B \frac{3}{2 m}\ln\left(\frac{3 h^2}{4 \pi  m^2}\frac{1}{\alpha(E,N ,\nu,\Omega )}\right) - \frac{k_B}{m} \ln \frac{\barn}{ m} -\frac{3}{2}\frac{k_B}{  m^3 \alpha(E,N ,\nu,\Omega )}\nu \barn.
\end{equation}
Equation \eqref{eq.Nstar} thus becomes 
\begin{equation}\label{eq.n.integral}
 \ln \rho +\frac{3}{2}\frac{1}{ m^2\alpha} \int d\rr' V(|\rr-\rr'|) \rho(\rr') = \ln\barn +\frac{3}{2}\frac{1}{ m^2\alpha}\nu \barn,
\end{equation}
which is a nonlinear integral equation for $\rho(\rr)$. 

Equation \eqref{eq.n.integral} can be linearized around $\barn$ through $\tilden := \frac{\rho}{\barn} -1\ll 1$ in magnitude
, which yields 
\begin{equation}\label{eq.n.integral.lin}
\tilden +\frac{3}{2}\frac{1}{ m^2\alpha} \barn\int d\rr' V(|\rr-\rr'|) \tilden(\rr') = 0, 
\end{equation}
which is a Fredholm integral equation of the second kind. This equation has a unique solution, which is homogeneous zero, if $|\Omega|<\infty$ and $\frac{3}{2}\frac{\barn}{ m^2\alpha}$ is not an eigenvalue of the integral operator, i.e. if $\max_{\rr \times\rr'\in \Omega \times \Omega} |V(|\rr-\rr'|)|=\max_{\rr\in \Omega} |V(|\rr|)|<\frac{2}{3}\frac{ m^2\alpha}{M}$.\footnote{Note that for $V(|\rr|)=|\rr|^k$, $k$ has to be negative.}
 

We have shown using physical (not mathematical) tools that there is a homogeneous solution to the macroscopic problem which is a MaxEnt equilibrium (maximizes the reduced entropy) and moreover entropy production is non-negative until this homogeneous density distribution is reached. Therefore, this approach is providing thermodynamic arguments for linear Landau damping.

\subsubsection{Some qualitative insight into macroscopic evolution equations. Linearization.}
It was shown in the preceding section that homogeneous equilibrium is a state ("the" state when near to the equilibrium) for which the mechanical equilibrium entropy $S^{(ME)}$ attains its maximum.  The approach of $\rho$ and $\eps$ by means of evolution equations \eqref{eq.Macro.Final} towards the homogeneous equilibrium can be thus expected.

We can gain some analytical insight into the governing macroscopic evolution equations from their analytical solution which is available only under certain assumptions. We shall thus solve linearized version of the equations around constant solutions.

The macroscopic evolution equations \eqref{eq.Macro.Final} can posses a constant (both in time and space) solution $\bar{\epsilon},\bar{\rho}$ only when $ \FF^{(ME)}=0=\bar{\rho} \int d\rr' \frac{\partial V(|\rr''-\rr'|)}{\partial r_j''}$, which requires the system $\Omega$ to be periodic as discussed above. Then actually arbitrary constant $\bar{\epsilon},\bar{\rho}$ is a solution but the thermodynamically admissible single one is chosen by the entropy functional as described above.

Denoting small perturbations around this constant solution with tildes, we may approximate the evolution equations to the leading (linear) order as follows ($ \tilde{\bar{\FF}}^{(ME)}$ stands for $ \tilde{\bar{F_j}}^{(ME)} = -\frac{1}{m^2}\int d\rr' \frac{\partial V(|\rr''-\rr'|)}{\partial r_j''} (\bar \rho(\rr')+\tilde\rho(\rr'))$
\begin{subequations} \label{eq.Macro.Lin}
\begin{eqnarray}
\frac{\partial \tilde{\rho}}{\partial t} &=& \frac{\tau}{2} \left[\frac 2 3  \Delta (\bar\eps+\tilde\eps) - \nabla\cdot((\bar\rho+\tilde\rho) \tilde{\bar{\FF}}^{(ME)})\right]\approx\nonumber\\
	&\approx&\frac{\tau}{2} \left[\frac{2}{3} \Delta\tilde\eps- \frac{\bar\rho}{m^2}  \int d\rr'\partial^2_j V(|\rr-\rr'|)\tilde \rho(\rr') \right],\\
\frac{\partial \tilde{\eps}}{\partial t} &=& \frac{\tau}{2} \left[\frac{10}{9}  \Delta\left((\bar\rho+\tilde\rho) \left(\frac{(\bar\eps+\tilde\eps)}{(\bar\rho+\tilde\rho)}\right)^2\right)  - \nabla\cdot((\bar\eps+\tilde\eps) \tilde{\bar{\FF}}^{(ME)}) +  (\bar\rho+\tilde\rho) \left(\tilde{\bar{\FF}}^{(ME)}\right)^2	\right]\approx\nonumber\\
&\approx& \frac{\tau}{2} \left[-\frac{10}{9} \left(\frac{\bar\eps}{\bar\rho}\right)^2 \Delta\tilde\rho+\frac{20}{9} \frac{\bar\eps}{\bar\rho} \Delta \tilde \eps-  \frac{\bar\eps}{m^2} \int d\rr'\partial^2_j V(|\rr-\rr'|)\tilde \rho(\rr') \right].
\end{eqnarray}
\end{subequations}

Rescaling time $\tilde t = \frac{\tau}{2} t$, dropping tildes and applying Fourier transform in space ($\rr\rightarrow\boldsymbol\xi$ and denoting the images with hat) yields
\begin{subequations} \label{eq.Macro.Lin.FT}
\begin{eqnarray}
\frac{\partial \hat{\rho}}{\partial t} &=&  - \boldsymbol\xi^2 \left[-\bar\rho  \frac{\hat V(\boldsymbol\xi)}{m^2}\hat \rho + \frac{2}{3} \hat\eps\right] ,\\
\frac{\partial \hat{\eps}}{\partial t} &=& - \boldsymbol\xi^2 \left[-\left(\frac{10}{9}  \left(\frac{\bar\eps}{\bar\rho}\right)^2+ \bar\eps  \frac{\hat V(\boldsymbol\xi)}{m^2}\right) \hat\rho+\frac{20}{9}\frac{\bar\eps}{\bar\rho} \hat \eps  \right],
\end{eqnarray}
\end{subequations}
where $\hat V$ is a Fourier transform of $V(|\rr|)$. For fixed $\xi_j$ we have a linear system of ODEs whose long-time behavior is governed by eigenvalues 
\begin{equation*}
	\lambda_\pm (\boldsymbol\xi)=-\boldsymbol\xi^2\left(\frac{20 \bar \eps-9\bar{\rho}^2 \hat V(\boldsymbol\xi)/ m^2}{18 \bar \rho}\pm\frac{1}{18 \bar \rho} \sqrt{160  \bar\eps^2+ 576 \bar\eps\bar \rho^2 \hat V(\boldsymbol\xi) /m^2+ 81 \bar\rho^4 \hat V^2(\boldsymbol\xi) /m^4}\right)
\end{equation*}
and the system \eqref{eq.Macro.Lin.FT} has a solution in terms of a linear combination of two exponentials $\exp({\lambda_\pm t})$. To explore the decay or growth of the perturbations $\rho,\eps$ we need to calculate inverse Fourier transform. Due to linearity of Fourier transform the decay of perturbations occurs iff $\mathcal{F}^{-1}\left[e^{-\lambda_\pm(\boldsymbol \xi) t}\right](t,\rr)$ decay with time. With known potential $V$ we may readily assess this condition. To illustrate this we further assume that the dependence of $\lambda_\pm$ on $\hat V(\boldsymbol\xi)$ can be dropped, i.e. $\bar\eps\gg\bar\rho^2\hat V(\boldsymbol\xi)$, and hence $\lambda_\pm\propto -\boldsymbol\xi^2$. Thence $\eps,\rho$ are linear combination of two fundamental solutions to diffusion operator. Particularly we have
\begin{equation}
	\eps\approx C_+ \frac{1}{(4\pi \bar\lambda_+ t)^{3/2}}\exp\left(-\frac{\rr^2}{4\bar\lambda_+ t}\right) + C_- \frac{1}{(4\pi \bar\lambda_- t)^{3/2}}\exp\left(-\frac{\rr^2}{4\bar\lambda_- t}\right)
\end{equation}
with $\bar\lambda_\pm=\frac{2\bar \eps}{9 \bar \rho}(5\pm\sqrt{10})$. Therefore the initial delta pulse (at the origin) is flattening out with time as $(t \frac{\tau}{2m})^{-3/2}$ which can be used for identification of the essential parameter in the Ehrenfest reduction, the time-scale $\tau$. 

Experiments measuring the Landau damping report characteristic times of these decays, hence assume exponential decay. Note, however, that $C_1 t^{-3/2}+C_2$ is practically indistinguishable for exponential decay over finite time intervals and real data. Finally note one should convolve the initial condition with the above identified responses to delta pulses but we argue that such a response is already providing a qualitative insight into long-time behavior (decay).

In summary, linearization of equations \eqref{eq.Macro.Final} around homogeneous solutions leads to the conclusion that the equations approach the homogeneous solutions, which is compatible with maximization of entropy at the homogeneous solutions. Landau damping can be thus manifestly visible on the less detailed level of mechanical equilibrium.


\section{Conclusion}
Landau damping is the property of the distribution function, governed by reversible Vlasov equations, that it approaches spatially homogeneous equilibria weakly. Particle density (integral of the distribution function) then approaches the homogeneous equilibrium strongly. The reversible Vlasov equation does not alter the Boltzmann entropy, which depends on the distribution function. Conservation of entropy and the approach to the homogeneous equilibrium (Landau damping) are thus seemingly in contrast.

Here we have shown (by means of the Ehrenfest reduction) that the Vlasov equation can be approximated by evolution equations for particle density $\rho$ and  kinetic energy density $\eps$, \eqref{eq.Macro.Final}. These state variables are equipped with their own entropy, \eqref{eq.S.ME}, which is given by maximization of the Boltzmann entropy. The evolution equations approach homogeneous equilibrium, where the reduced entropy is maximized. It is thus the reduced hydrodynamic entropy \eqref{eq.S.ME} that is maximized during the Landau damping, and Landau damping has been given an alternative thermodynamic interpretation.

\section*{Acknowledgment}
This work was supported by Czech Science Foundation, project no.  17-15498Y, and by Natural Sciences and Engineering Research Council of Canada (NSERC).



\end{document}